\documentclass[12pt]{article}
\usepackage{epsf}
\usepackage{amsmath, amssymb}

\begin{document}

\newcommand{\rem}[1]{{\bf #1}}

\renewcommand{\thefootnote}{\fnsymbol{footnote}}
\setcounter{footnote}{0}
\begin{titlepage}

\def\thefootnote{\fnsymbol{footnote}}

\begin{center}

\hfill UT-10-12\\
\hfill July, 2010\\

\vskip .75in

{\Large \bf 
Spin and Chirality Determination of Superparticles
\\
with Long-Lived Stau at the LHC
}

\vskip .75in

{\large
Takumi Ito$^{(a,b)}$ and Takeo Moroi$^{(a,c)}$
}

\vskip 0.25in

$^{(a)}${\em
Department of Physics, University of Tokyo,
Tokyo 113-0033, JAPAN}

\vskip 0.07in

$^{(b)}${\em
Department of Physics, Tohoku University,
Sendai 980-8578, JAPAN}

\vskip 0.07in
$^{(c)}${\em
Institute for the Physics and Mathematics of the Universe,\\
University of Tokyo, Chiba 277-8568, Japan}

\end{center}
\vskip .5in

\begin{abstract}

  Long-lived stau shows up in various supersymmetric models, like
  gauge-mediated supersymmetry (SUSY) breaking model.  At the LHC
  experiment, long-lived stau is useful not only for the discovery of
  SUSY signals but also for the study of the detailed properties of
  superparticles.  We discuss a method to obtain information on spins
  and chiralities of superparticles in the framework with long-lived
  stau.  We also show that such a study can be used to distinguish
  SUSY model from other models of new physics, like the universal
  extra dimension model.

\end{abstract}

\end{titlepage}

\renewcommand{\thepage}{\arabic{page}}
\setcounter{page}{1}
\renewcommand{\thefootnote}{\#\arabic{footnote}}
\setcounter{footnote}{0}

Low energy supersymmetry (SUSY) is an attractive candidate of the
physics beyond the standard model, and superparticles are important
targets of the LHC experiment.  Even if signals of superparticles are
found, however, it is non-trivial to confirm that the newly discovered
particles are superparticles.  This is because the SUSY-like mass
spectrum is possible in some class of models other than supersymmetric
one.  Thus, once exotic particles are found at the LHC, their
properties should be studied in detail to understand the underlying
model.

Procedure to study the properties of superparticles crucially depends
on their mass spectrum; in particular, for each candidate of the
lightest supersymmetric particle (LSP) in the minimal supersymmetric
standard model (MSSM) sector, which we call MSSM-LSP, we expect
different type of SUSY signals at the LHC experiment.  Even though the
lightest neutralino is the most popular candidate of the MSSM-LSP,
charged (and/or colored) superparticle can also be the MSSM-LSP if it
is unstable.  Thus, for each candidate of the MSSM-LSP, we should
consider how and how well the SUSY events can be studied at the LHC.

In the present study, we consider an important possibility that the
lighter stau $\tilde{\tau}$ is the MSSM-LSP.  Stau can be the MSSM-LSP
in well-motivated SUSY breaking scenarios, like the gauge mediated
SUSY breaking scenario \cite{Dine:1995ag, Dine:1994vc}.  Even though
$\tilde{\tau}$ is expected to be unstable in such a case, its decay
length can be much longer than the size of the LHC detectors ($\sim
10\ {\rm m}$).  Then, $\tilde{\tau}$ is regarded as a long-lived
charged particle at the LHC experiment.  For example, in the
gauge-mediation model, $\tilde{\tau}$ decays into $\tau$ + gravitino,
and the decay length becomes longer than $10\ {\rm m}$ if the
gravitino mass is heavier than $\sim 1\ {\rm keV}$.  Then, in such a
case, we can observe a track of $\tilde{\tau}$ at the LHC experiment
and we expect very unique signals at the LHC.  The $\tilde{\tau}$
track should be useful not only for the discovery but also for the
study of the properties of superparticle \cite{Hinchliffe:1998ys,
  Ambrosanio:2000ik, Hamaguchi:2004df, Feng:2004yi, Buchmuller:2004rq,
  Ellis:2006vu, Ibe:2007km, Rajaraman:2007ae, Ishiwata:2008tp,
  Kaneko:2008re, Kitano:2008sa, Asai:2009ka, Feng:2009yq,
  Biswas:2009zp, Ito:2009xy, Biswas:2009rba, Kitano:2010tt}; in
particular, in the previous study, we have shown that the masses of
squark, sleptons, and neutralinos can be measured with very small
errors using long-lived $\tilde{\tau}$ \cite{Ito:2009xy}.

In this letter, we extend our previous analysis to discuss how we can
study the properties, in particular, spins and chiralities (i.e.,
handednesses), of the superparticles in the decay chain.  Spin and
chirality measurements for the neutralino-MSSM-LSP case have been
discussed in many literatures \cite{Barr:2004ze, Smillie:2005ar,
  Datta:2005zs, Choi:2006mt, Wang:2006hk, Alves:2007xt, Csaki:2007xm,
  Kane:2008kw, Ehrenfeld:2009rt}. We will see that, if $\tilde{\tau}$
is the MSSM-LSP, the study becomes easier because SUSY events can be
distinguished from standard-model backgrounds by identifying
$\tilde{\tau}$-track, and also because full event reconstruction is
possible due to the absence of the missing momentum.  Information on
the particles in the decay chain is extracted from the invariant-mass
distribution of the decay products.  Unfortunately, the observed
invariant-mass distributions are deformed from the parton-level
predictions, which becomes the origin of systematic uncertainties in
the test of underlying models.  We propose to analyze the ratio of the
numbers of two different processes with same event topology, from
which many of the systematic uncertainties should cancel out.  We will
see that such an analysis is useful to understand the chiralities of
the superpaticles in the decay chain.  Furthermore, we can also
distinguish SUSY model from other models with particles with different
spin, like the universal extra dimension (UED) model
\cite{Appelquist:2000nn}, which may result in a SUSY-like mass
spectrum.  We note here that, in order to make our points clear, we
consider supersymmetric standard model with long-lived $\tilde{\tau}$.
However, the procedure should work in other class of models, like the
UED model with a long-lived Kalza-Klein (KK) charged lepton.\footnote
{In the simplest UED model, KK mode of the $U(1)_Y$ is the lightest KK
  particle (LKP).  However, such a mass spectrum can be easily
  modified by introducing the brane-localized interactions.  Then, the
  long-lived KK lepton may show up when the KK mode of the graviton is
  the LKP while the KK mode of a lepton is the second-lightest KK
  particle.}

Let us start with discussing the underlying model we have in mind,
i.e., supersymmetric model with $\tilde{\tau}$-MSSM-LSP.  If
low-energy SUSY exists, squarks and gluino are particularly produced
at the LHC.  In the following discussion, we consider the case where
the gluino is heavier than squarks.  (Such a mass spectrum is
naturally realized, for example, in gauge mediation model with
$\tilde{\tau}$-MSSM-LSP.)  In such a case, the processes
$pp\rightarrow\tilde{u}\tilde{u}$ and $\tilde{d}\tilde{d}$ have
significant cross section compared to other processes because of large
parton densities of the up- and down-quarks in proton.  (Other types
of SUSY processes also occur, but their effects become subdominant in
the following study by imposing relevant kinematical cuts.)  Then,
once $\tilde{q}$ (where, here and hereafter, $q$ is for $u$ and $d$)
is produced, it may cause the decay chain $\tilde{q}\rightarrow
q\chi^0_1$, followed by $\chi^0_1\rightarrow\tau^\pm\tilde{\tau}^\mp$
(with $\chi^0_1$ being the lightest neutralino); for simplicity, we
denote such a decay chain as $\tilde{q}\rightarrow
q\tau^\pm\tilde{\tau}^\mp$.  In the following, we only use the
hadronic decay mode of $\tau$.  Then, SUSY events result in
$\tilde{\tau}$ tracks and $\tau$-jets as well as energetic jets. If
the velocity of $\tilde{\tau}$ is small enough, SUSY events can be
distinguished from the standard-model events by identifying
$\tilde{\tau}$ track.  The discovery of such events should be a clear
indication of the existence of a new physics beyond the standard
model.  In addition, with the study of the endpoints or peak positions
of the invariant-mass distributions, information on the masses of the
new particles (i.e., squarks and neutralinos) will be obtained
\cite{Ito:2009xy}.

Once the masses of newly discovered particles are determined, the next
task will be to precisely understand the underlying model.  In
particular at the early stage of the LHC experiment, it is non-trivial
to confirm that the underlying model is MSSM.  In the present case,
the existence of particles with masses of $m_{\tilde{q}_R}$ and
$m_{\chi^0_1}$ as well as a long-lived charged particle can be experimentally
confirmed.  However, if the masses are the only information available
from the experiment, it is not clear if those particles are
superparticles.  In addition, even if the underlying model is assumed
to be the MSSM, chiralities of observed $\tilde{q}$ and $\tilde{\tau}$
are unknown.  As we have mentioned, the information on the underlying
model is imprinted in the invariant-mass distributions of the
particles from the decay; in order to confirm or exclude a specific
underlying model, one should check the consistency between observed
invariant-mass distribution and prediction of the postulated model.
In the following, we discuss how and how well we can perform such an
analysis.

In the present case, invariant mass of $(q,\tau)$ system contains
important information.  For the study of the SUSY model, we
parameterize the relevant interaction terms as
\begin{eqnarray}
  {\cal L}_{\rm int} = 
  \bar{\chi}^0_1 (g_{q,L} P_L + g_{q,R} P_R ) q \tilde{q}^*
  +
  \bar{\chi}^0_1 (g_{\tau,L} P_L + g_{\tau,R} P_R )  \tau \tilde{\tau}^*
  + {\rm h.c.},
\end{eqnarray}
where $g_{q,L}$, $g_{q,R}$, $g_{\tau,L}$, and $g_{\tau,R}$ are
coupling constants.  (In the following, we consider the case that the
production of right-handed squarks plays an important role and also
that the lighter stau is right-handed.  Then, for the process we will
study, $g_{q,L}=g_{\tau,L}=0$.)  The invariant-mass distributions of the
decay processes are given by
\begin{eqnarray}
  \frac{1}{2\Gamma_{\tilde{q}\rightarrow q\tau^+\tilde{\tau}^-} }
  \frac{d\Gamma_{\tilde{q}\rightarrow q\tau^+\tilde{\tau}^-}}{d x_{q\tau}}
  &=&
  \frac{(g_{q,L}^2 g_{\tau,L}^2 + g_{q,R}^2 g_{\tau,R}^2 ) (1 - x_{q\tau}) 
    + (g_{q,L}^2 g_{\tau,R}^2 + g_{q,R}^2 g_{\tau,L}^2 )  x_{q\tau}}
  {g_{q,L}^2 g_{\tau,L}^2 + g_{q,R}^2 g_{\tau,R}^2
    + g_{q,L}^2 g_{\tau,R}^2 + g_{q,R}^2 g_{\tau,L}^2},
  \label{dG/dx(tau+)}
  \\
  \frac{1}{2\Gamma_{\tilde{q}\rightarrow q\tau^-\tilde{\tau}^+}}
  \frac{d\Gamma_{\tilde{q}\rightarrow q\tau^-\tilde{\tau}^+}}{d x_{q\tau}}
  &=&
  \frac{(g_{q,L}^2 g_{\tau,L}^2 + g_{q,R}^2 g_{\tau,R}^2 ) x_{q\tau}
    + (g_{q,L}^2 g_{\tau,R}^2 + g_{q,R}^2 g_{\tau,L}^2 ) (1 - x_{q\tau})}
  {g_{q,L}^2 g_{\tau,L}^2 + g_{q,R}^2 g_{\tau,R}^2
    + g_{q,L}^2 g_{\tau,R}^2 + g_{q,R}^2 g_{\tau,L}^2},
  \label{dG/dx(tau-)}
\end{eqnarray}
and $d\Gamma_{\tilde{q}^*\rightarrow
  q\tau^\pm\tilde{\tau}^\mp}/dx_{q\tau}=d\Gamma_{\tilde{q}\rightarrow
  q\tau^\mp\tilde{\tau}^\pm}/dx_{q\tau}$, where
\begin{eqnarray}
  x_{q\tau} = M_{q\tau}^2 / \hat{M}_{q\tau}^2 
  = (p_q + p_\tau)^2 / \hat{M}_{q\tau}^2,
\end{eqnarray}
with $\hat{M}_{q\tau}^2$ being the maximal value of $M_{q\tau}^2$:
\begin{eqnarray}
  \hat{M}_{q\tau}^2 = 
  \frac{(m_{\tilde{q}}^2 - m_{\chi^0_1}^2)(m_{\chi^0_1}^2-m_{\tilde{\tau}}^2)}
  {m_{\chi^0_1}^2}.
\end{eqnarray}
Because of the Majorana nature of $\chi^0_1$,
$\Gamma_{\tilde{q}\rightarrow
  q\tau^+\tilde{\tau}^-}=\Gamma_{\tilde{q}\rightarrow
  q\tau^-\tilde{\tau}^+}$.  One can easily see that
$d\Gamma_{\tilde{q}\rightarrow q\tau^\pm\tilde{\tau}^\mp}/dx_{q\tau}$
has non-trivial dependence on $x_{q\tau}$.  We also note here that the
distributions of $M_{q\tau}^2$ depend on the chiralities of
$\tilde{q}$ and $\tilde{\tau}$ and that the invariant-mass
distributions are different for $\tilde{q}\rightarrow
q\tau^+\tilde{\tau}^-$ and $\tilde{q}\rightarrow
q\tau^-\tilde{\tau}^+$.  These facts are important in the following
discussion.  The distributions given in Eqs.\ \eqref{dG/dx(tau+)} and
\eqref{dG/dx(tau-)} are crucial check points of the present model.

Now, we show how the observed invariant-mass distributions behave by
using the MC analysis.  In our study, we work in the framework of
gauge-mediated model.  We consider the situation that the MSSM-LSP is
lighter stau $\tilde{\tau}$, which is assumed to be long-lived, and
that the processes $pp\rightarrow\tilde{u}\tilde{u}$ and
$\tilde{d}\tilde{d}$ have large cross sections.  We adopt the
following parameters:
\begin{eqnarray}
  \Lambda = 60\ {\rm TeV}, \ \ 
  M_{\rm mess} = 900\ {\rm TeV}, \ \ 
  N_{\bf 5} = 3,\ \ 
  \tan\beta = 35,\ \ 
  {\rm sign} (\mu) = +,
\end{eqnarray}
where $\Lambda$ is the ratio of the $F$- and $A$-components of the
SUSY breaking field, $M_{\rm mess}$ is the messenger scale, $N_{\bf
  5}$ is the number of messenger multiplets in units of ${\bf 5}+{\bf
  \bar{5}}$ representation of $SU(5)$ grand-unified group, $\tan\beta$
is the ratio of the vacuum expectation values of two Higgs bosons, and
$\mu$ is the SUSY invariant Higgs mass.  The mass spectrum of
superparticles is calculated by using ISAJET 7.64~\cite{Paige:2003mg};
the result is summarized in Table~\ref{table:susymass}.  The LHC
phenomenology of this parameter point has been studied in
\cite{Ito:2009xy}, which has shown that the masses of superparticles
can be determined with relatively small uncertainties.  In particular,
$m_{\tilde{q}_R}$ and $m_{\chi^0_1}$ are measured with the accuracies
of $\sim 10\ {\rm GeV}$ and $\sim 1\ {\rm GeV}$, respectively, with
the luminosity of ${\cal L}=100\ {\rm fb}^{-1}$.  In addition,
$m_{\tilde{\tau}}$ can be also determined by combining time-of-flight
and momentum information; the expected accuracy is $\sim 0.1\ {\rm
  GeV}$ \cite{Ellis:2006vu}.  In our study, we assume that the masses
of these superparticles can be well determined before the study of the
invariant-mass distributions.

\begin{table}
  \centering
  \begin{tabular}{cr} 
    \hline \hline
    Particle & Mass (GeV) \\
    \hline
    $\tilde{g}$      & 1309.39  \\
    $\tilde{u}_{L}$  & 1231.70   \\
    $\tilde{u}_{R}$  & 1183.97   \\
    $\tilde{d}_{L}$  & 1234.28   \\
    $\tilde{d}_{R}$  & 1180.19   \\
    $\tilde{t}_{1}$  & 1082.85 \\
    $\tilde{t}_{2}$  & 1195.08 \\
    $\tilde{b}_{1}$  & 1145.24 \\
    $\tilde{b}_{2}$  & 1185.83 \\
    $\tilde{\nu_l}$  & 388.05 \\
    $\tilde{l}_{L}$  & 396.19 \\
    $\tilde{\tau}_{2}$ & 402.57 \\
    $\tilde{\nu}_\tau$ &  383.80 \\
    $\tilde{l}_{R}$  & 194.39 \\
    $\tilde{\tau}_{1}$ & 148.83 \\
    $\chi^0_{1}$ & 239.52 \\
    $\chi^0_{2}$ & 425.92 \\
    $\chi^0_{3}$ & 508.41 \\
    $\chi^0_{4}$ & 548.67 \\
    $\chi^\pm_{1}$ & 425.45 \\ 
    $\chi^\pm_{2}$ & 548.43 \\
    $h$ & 115.01 \\ 
    \hline\hline
  \end{tabular}
  \caption{\small Masses of the superparticles and the lightest 
    Higgs boson $h$ in units of GeV.  The input  parameters are
    $\Lambda = 60\ {\rm TeV}$, $M_{\rm mess} = 900\ {\rm TeV}$, 
    $N_{\bf 5} = 3$, 
    $\tan\beta = 35$, ${\rm sign} (\mu) = +$.
    (We use the top-quark mass of $171.3\ {\rm GeV}$.)}
  \label{table:susymass}
\end{table}

We have generated SUSY events for $\sqrt{s}=14\ {\rm TeV}$ with HERWIG
6.510 package \cite{Corcella:2000bw,Moretti:2002eu}. (The total cross
section of the SUSY events is $669.6\ {\rm fb}$.)  In order to
simulate detector effects, we use the PGS4 detector simulator
\cite{PGS4} with slight modification to treat stable stau; the
momentum resolution of $\tilde{\tau}$ is assumed to be the same as
those of muons.  Following \cite{Ambrosanio:2000ik}, we assume that
$\tilde{\tau}$ with $0.4\leq\beta_{\tilde{\tau}}\leq 0.91$ can be
detected with the efficiency of 100\ \% with no standard-model
background.  Staus with $\beta_{\tilde{\tau}}\geq 0.91$ are assumed to
be identified as muons.

In order to use the events with the decay chain $\tilde{q}\rightarrow
q\chi^0_1$, followed by $\chi^0_1\rightarrow\tau^\pm\tilde{\tau}^\mp$,
the following selection cuts are applied:
\begin{itemize}
\item[(a)] At least one $\tilde{\tau}$ with the velocity
  $0.4\leq\beta_{\tilde{\tau}}\leq 0.91$.
\item[(b)] At least one $\tau$-tagged jet $j_\tau$ with $p_T > 20\
  {\rm GeV}$.
\item[(c)] Exactly two jets $j$ with $p_T > 30\ {\rm GeV}$.
\end{itemize}
The requirement (c) is to eliminate the gluino production events.  If
there exists only one $\tilde{\tau}$ with
$0.4\leq\beta_{\tilde{\tau}}\leq 0.91$, the highest $p_T$ muon-like
object is regarded as second $\tilde{\tau}$ because two staus are
expected in SUSY events.  Then, for all the possible combinations of
$(j, j_\tau, \tilde{\tau})$, we perform the following study.  We first
reconstruct the tau momentum $p_\tau$ assuming that tau and stau are
from the decay of $\chi^0_1$ (whose mass is expected to be already
known).  Because the tau from the neutralino decay is highly boosted,
we approximate that the three-momentum of $\tau$ is parallel to that
of $j_\tau$.  Then, we obtain
\begin{eqnarray}
  p_\tau = z_{\tilde{q}}^{-1} p_{j_\tau},
\end{eqnarray}
where
\begin{eqnarray}
  z_{\tilde{q}} = 
  \frac{2p_{j_\tau} p_{\tilde{\tau}}}
  {m_{\chi^0_1}^2 - m_{\tilde{\tau}}^2}.
  \label{zqdef}
\end{eqnarray}
Combinations with $z_{\tilde{q}}>1$ is eliminated.  Then, we calculate
\begin{eqnarray}
  M_{\tilde{q}}=\sqrt{(p_{j}+p_{\tilde{\tau}}+z_{\tilde{q}}^{-1}p_{j_\tau})^2}.
\end{eqnarray}
Using the fact that there exists a very sharp peak in the distribution
of $M_{\tilde{q}}$, which is from $\tilde{q}_R$ production, only the
combinations with $m_{\tilde{q}_R}^{\rm (peak)}-40\ {\rm GeV}\leq
M_{\tilde{q}}\leq m_{\tilde{q}_R}^{\rm (peak)}+40\ {\rm GeV}$ are
adopted, where $m_{\tilde{q}_R}^{\rm (peak)}=1170\ {\rm GeV}$ is the
position of the peak \cite{Ito:2009xy}.\footnote
{The peak position in the present study is found to be smaller than
  the input value of the squark mass by $\sim 10\ {\rm GeV}$.  This is
  expected to be due to the energy leakage in the jet reconstruction,
  and may be corrected once the jet energy is well calibrated.}
We calculate the distribution of the following variable:
\begin{eqnarray}
  x_{j\tau} 
  \equiv M_{j\tau}^2 / \hat{M}_{j\tau}^2
  \equiv (p_j + z_{\tilde{q}}^{-1}p_{j_\tau})^2 / \hat{M}_{j\tau}^2,
\end{eqnarray}
where
\begin{eqnarray}
  \hat{M}_{j\tau}^2 = 
  \frac{(M_{\tilde{q}}^2 - m_{\chi^0_1}^2)(m_{\chi^0_1}^2-m_{\tilde{\tau}}^2)}
  {m_{\chi^0_1}^2}.
\end{eqnarray}
The charges of $j_\tau$ and $\tilde{\tau}$, which are both observable,
should be opposite for signal events.

\begin{figure}[t]
\begin{tabular}{cc}
  \begin{minipage}{0.5\hsize}
    \centerline{\epsfxsize=\textwidth\epsfbox{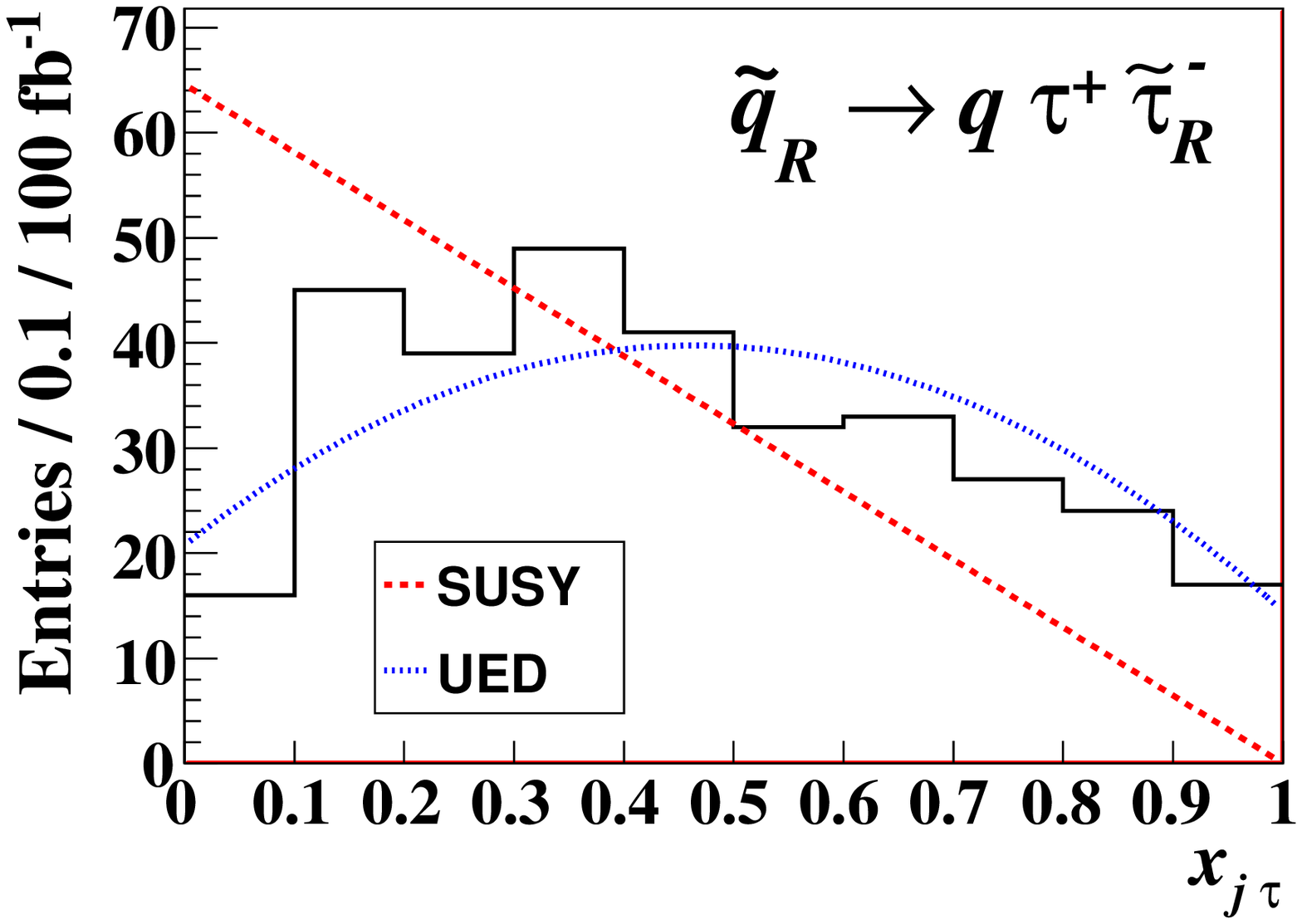}}
  \end{minipage}
  \begin{minipage}{0.5\hsize}
    \centerline{\epsfxsize=\textwidth\epsfbox{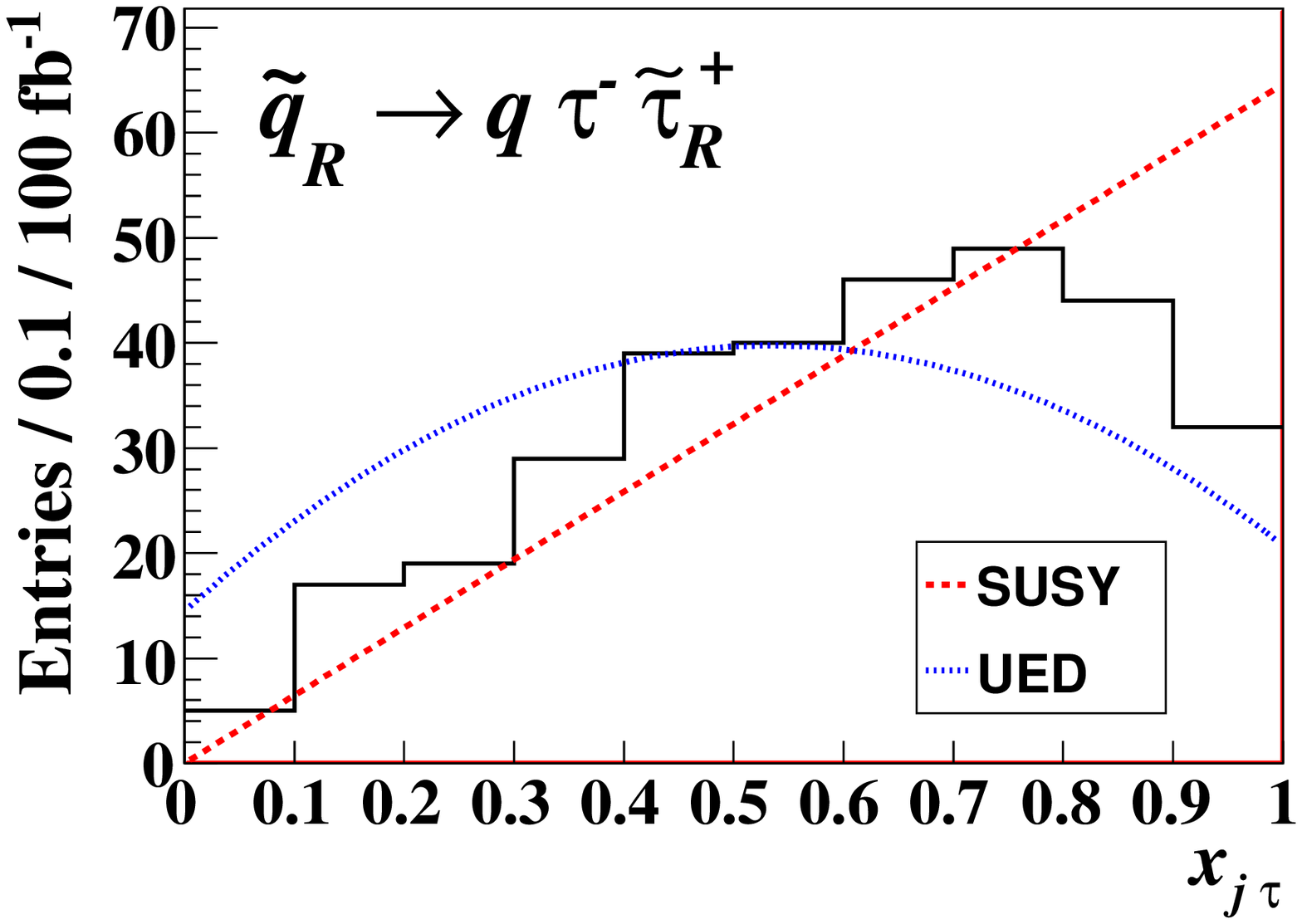}}
  \end{minipage}
\end{tabular}
\caption{\small Distributions of $x_{j \tau}$ in the reconstructed
  decay processes, $\tilde{q}_{R}\to q \tau^{+} \tilde{\tau}_{R}^{-}$
  (left) and $\tilde{q}_{R}\to q \tau^{-} \tilde{\tau}_{R}^{+}$
  (right), where $x_{j \tau}$ is a squared invariant mass normalized
  by its maximal value.  Here we take ${\cal L}=100\ {\rm fb}^{-1}$.
  The theoretical predictions are also shown in the SUSY case (red
  dashed line), and for similar processes in the UED case (blue dotted
  line).}
  \label{fig:hist}
\end{figure}

In Fig.\ \ref{fig:hist}, we show the distributions of
$x_{j\tau}=M_{j\tau}^2/\hat{M}_{j\tau}^2$ for $(j, j_\tau^+,
\tilde{\tau}^-)$ and $(j, j_\tau^-, \tilde{\tau}^+)$ events.  In the
same figure, we also plot the theoretical differential decay rates
given in Eqs.\ \eqref{dG/dx(tau+)} and \eqref{dG/dx(tau-)}; the
normalization is determined so that the total number of events agrees
with the MC data.  The MC results and the theoretical predictions seem
to agree at qualitative level.

At quantitative level, however, the agreement is not perfect.  We can
see that the numbers of events are suppressed when
$x_{j\tau}\rightarrow 0$ and $1$.  These can be understood by
considering the event configurations in those limits.  When
$x_{j\tau}\rightarrow 0$, $\tau$ is emitted in the opposite direction
to $\chi^0_1$ in the rest frame of $\tilde{q}$.  Because most of the
squarks are not significantly boosted with the present choice of the
squark masses, $\tau$ in events with small $x_{j\tau}$ are likely to
have small $p_T$.  Because the $p_T$ of $j_\tau$ is required to be
larger than $20\ {\rm GeV}$, the acceptance of the signal is
suppressed for $x_{j\tau}\sim 0$.  On the contrary, when
$x_{j\tau}\rightarrow 1$, the momenta of $\tau$ and $\tilde{\tau}$
become parallel in the rest frame of $\tilde{q}$.  Such events are
eliminated by the isolation cut for the $\tau$-jet; in the present
analysis, no extra activity is allowed in the cone (with the size of
$\Delta R=0.5$) around $\tau$-jet.  To see the validity of these
arguments, we estimate the efficiencies of corresponding kinematical
cuts using the momentum information on the decay products obtained
from HERWIG output.  The results are shown in Tables
\ref{table:eff_pT} and \ref{table:eff_dR}.  We can see that the
efficiencies behave as we have expected.  It is also notable that the
efficiencies depend on $x_{j\tau}$, but are insensitive to the charges
of final-state $\tau$ and $\tilde{\tau}$.

Another important reason of the disagreement should be the
contamination of background, which can be from fake $\tau$-jets
(mis-identified QCD jets) as well as from wrong combination where
$\tau$ and $\tilde{\tau}$ have different parents.  Once the real data
will become available, an accurate determination of the number of
backgrounds may be possible using, for example, events off from the
squark-mass peak (i.e., the sideband).  In the present MC analysis, we
have estimated the shape of the background from the sideband samples,
and the number of backgrounds in each bin is inferred to be
approximately universal as far as $x_{j\tau}$ is not close to $0$ or
$1$.  In the following analysis, we adopt constant background in each
bin.  Since the accurate estimation of the total number of backgrounds
is difficult from the sideband samples in the present analysis, the
normalization of background is treated as a free parameter and is
determined so that the $\chi^2$ variable defined below is minimized.

\begin{table}[t]
  \centering
  \begin{tabular}{ccc}
    \hline \hline
    Bin & $(j, j_\tau^+, \tilde{\tau}^-)$ & $(j, j_\tau^-, \tilde{\tau}^+)$
    \\
    \hline
    $0  \leq x_{q\tau} < 0.1$ & 0.09 & 0.13 \\
    $0.1\leq x_{q\tau} < 0.2$ & 0.19 & 0.21 \\
    $0.2\leq x_{q\tau} < 0.3$ & 0.24 & 0.24 \\
    $0.3\leq x_{q\tau} < 0.4$ & 0.25 & 0.25 \\
    $0.4\leq x_{q\tau} < 0.5$ & 0.25 & 0.26 \\
    $0.5\leq x_{q\tau} < 0.6$ & 0.27 & 0.27 \\
    $0.6\leq x_{q\tau} < 0.7$ & 0.27 & 0.26 \\
    $0.7\leq x_{q\tau} < 0.8$ & 0.29 & 0.28 \\
    $0.8\leq x_{q\tau} < 0.9$ & 0.28 & 0.28 \\
    $0.9\leq x_{q\tau} < 1.0$ & 0.32 & 0.28 \\
    \hline\hline
  \end{tabular}
  \caption{\small Ratio of the total number of squark-decay events
    and that with $p_T(j_\tau)\geq 20\ {\rm GeV}$.}
  \label{table:eff_pT} 
  \vspace{7mm}
  \centering
  \begin{tabular}{ccc}
    \hline \hline
    Bin & $(j, j_\tau^+, \tilde{\tau}^-)$ & $(j, j_\tau^-, \tilde{\tau}^+)$
    \\
    \hline
    $0  \leq x_{q\tau} < 0.1$ & 0.97 & 0.98 \\
    $0.1\leq x_{q\tau} < 0.2$ & 0.95 & 0.94 \\
    $0.2\leq x_{q\tau} < 0.3$ & 0.88 & 0.87 \\
    $0.3\leq x_{q\tau} < 0.4$ & 0.80 & 0.79 \\
    $0.4\leq x_{q\tau} < 0.5$ & 0.72 & 0.71 \\
    $0.5\leq x_{q\tau} < 0.6$ & 0.62 & 0.61 \\
    $0.6\leq x_{q\tau} < 0.7$ & 0.49 & 0.50 \\
    $0.7\leq x_{q\tau} < 0.8$ & 0.41 & 0.39 \\
    $0.8\leq x_{q\tau} < 0.9$ & 0.29 & 0.29 \\
    $0.9\leq x_{q\tau} < 1.0$ & 0.26 & 0.18 \\
    \hline\hline
  \end{tabular}
  \caption{\small Ratio of the total 
    number of squark-decay 
    event and that with
    $\Delta R_{j_\tau\tilde{\tau}}\geq 0.5$.}
  \label{table:eff_dR} 
\end{table}

If the effects of the deformation will be well understood in future
by, for example, a reliable MC analysis, the invariant-mass
distributions may be directly used to discriminate underlying models.
However, it is desirable to find quantities which are insensitive to
the effects of deformation.  For this purpose, we consider the ratio
of the numbers of $(j,j_\tau^+,\tilde{\tau}^-)$ and
$(j,j_\tau^-,\tilde{\tau}^+)$ events; we define
\begin{eqnarray}
  L_i = 
  \ln \frac{N_i(j,j_\tau^+,\tilde{\tau}^-)}
  {N_i(j,j_\tau^-,\tilde{\tau}^+)},
\end{eqnarray}
where $N_i(j,j_\tau^\pm,\tilde{\tau}^\mp)$ are the numbers of events
in $i$-th bin with charges of $(j_\tau, \tilde{\tau})$ being
$(\pm,\mp)$.  The error of $L_i$ is given by
\begin{eqnarray}
  \delta L_i^2 = 
  \frac{1}{N_i(j,j_\tau^+,\tilde{\tau}^-)}
  +\frac{1}{N_i(j,j_\tau^-,\tilde{\tau}^+)}.
\end{eqnarray}
To discuss how well the theoretical prediction is expected to agree
with experimental result, we calculate
\begin{eqnarray}
  \chi^2 \equiv \sum_i
  \frac{(L_i - L_i^{\rm (th)})^2}{\delta L_i^2},
\end{eqnarray}
where $L_i^{\rm (th)}$ is the theoretical prediction which is given by
\begin{eqnarray}
  L_i^{\rm (th)} = \ln 
  \frac{N_i^{\rm (signal)}(j,j_\tau^+,\tilde{\tau}^-) + N^{\rm (BG)}}
  {N_i^{\rm (signal)}(j,j_\tau^-,\tilde{\tau}^+) + N^{\rm (BG)}},
\end{eqnarray}
and the summation is over the bins in the range of $0.1\leq
x_{j\tau}\leq 0.9$ (with the width of $\Delta x_{j\tau}=0.1$); in
order to minimize the effects of background contamination, we do not
use the bins at $x_{j\tau}\sim 0$ and $1$.  Here, $N_i^{\rm
  (signal)}(j,j_\tau^\pm,\tilde{\tau}^\mp)$ are theoretical predictions
of the number of signal events in $i$-th bin calculated from Eqs.\
\eqref{dG/dx(tau+)} and \eqref{dG/dx(tau-)}, while $N^{\rm (BG)}$ is
the number of background events in each bin, which is independent of
$i$.

\begin{figure}[t]
  \centerline{\epsfxsize=0.75\textwidth\epsfbox{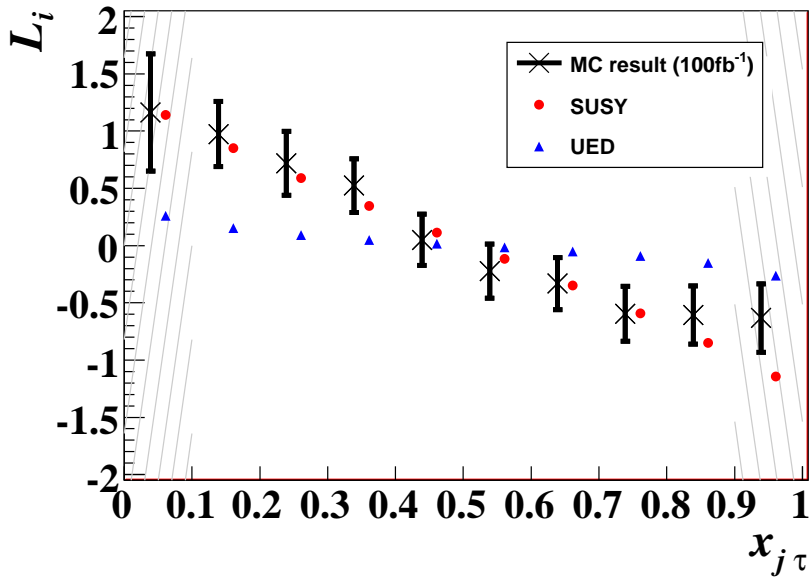}}
  \caption{\small Distribution of $L_{i}$.  MC result for $\mathcal{L}
    = 100$ fb$^{-1}$ is cross with error bar.  Theoretical predictions
    of $L_{i}$ obtained after minimizing $\chi^{2}$ are also shown for
    the SUSY case (red circles), and the UED case (blue triangles).  }
  \label{fig:ratio}
\end{figure}

$L_i$ from the MC analysis is shown in Fig.\ \ref{fig:ratio} for
${\cal L}=100\ {\rm fb}^{-1}$.  In the same figure, theoretical
predictions $L_i^{\rm (th)}$ are also shown.  We can see that the MC
result well agrees with theoretical prediction of the present SUSY
model.  In addition, in Table \ref{table:Li}, we show the result for
an ideal case where the luminosity is sufficiently large.  (Here, we
generate the event for ${\cal L}=20\ {\rm ab}^{-1}$.)  The results in
the table indicate that the effects of the deformation of the
invariant-mass distribution almost cancel out by taking the ratio.

\begin{table}[t]
  \centering
  \begin{tabular}{crrrr}
    \hline \hline
    Bin & $L_i$ & $\delta L_i^{(100)}$ & $L_i^{\rm (th, SUSY)}$ &  
    $L_i^{\rm (th, UED)}$
    \\
    \hline
    $0  \leq x_{j\tau} < 0.1$ &  $0.97$ & $0.45$ &  $0.91$ &  $0.26$ \\
    $0.1\leq x_{j\tau} < 0.2$ &  $0.72$ & $0.30$ &  $0.69$ &  $0.15$ \\
    $0.2\leq x_{j\tau} < 0.3$ &  $0.50$ & $0.27$ &  $0.48$ &  $0.09$ \\
    $0.3\leq x_{j\tau} < 0.4$ &  $0.29$ & $0.25$ &  $0.29$ &  $0.05$ \\
    $0.4\leq x_{j\tau} < 0.5$ &  $0.07$ & $0.24$ &  $0.09$ &  $0.02$ \\
    $0.5\leq x_{j\tau} < 0.6$ & $-0.12$ & $0.24$ & $-0.09$ & $-0.02$ \\
    $0.6\leq x_{j\tau} < 0.7$ & $-0.31$ & $0.24$ & $-0.29$ & $-0.05$ \\
    $0.7\leq x_{j\tau} < 0.8$ & $-0.47$ & $0.25$ & $-0.48$ & $-0.09$ \\
    $0.8\leq x_{j\tau} < 0.9$ & $-0.65$ & $0.26$ & $-0.69$ & $-0.15$ \\
    $0.9\leq x_{j\tau} < 1.0$ & $-0.86$ & $0.32$ & $-0.91$ & $-0.26$ \\
    \hline\hline
  \end{tabular}
  \caption{\small $L_i$, $\delta L_i^{(100)}$ (which is the error for
    the case of ${\cal L}=100\ {\rm fb}^{-1}$), and
    $L_i^{\rm (th)}$ for the case of large enough luminosity.
    The error of $L_i$ for the luminosity ${\cal L}$ is given by
    $\delta L_i=\delta L_i^{(100)}/\sqrt{{\cal L}_{100}}$ with
    ${\cal L}_{100}$ being the luminosity in units of
    $100\ {\rm fb}^{-1}$.}
  \label{table:Li}
\end{table}

The value of $\chi^2$ varies as we take different sets of event
samples.  In order to estimate the typical value of $\chi^2$, we
calculate $\chi^2$ for 20 sets of MC samples (for a fixed value of
luminosity ${\cal L}$), and obtain averaged value of $\chi^2$.  For
${\cal L}=30$ and $100\ {\rm fb}^{-1}$, the averaged values are found
to be $\langle\chi^2\rangle=7.6$ and $8.0$, respectively, which
indicates a good agreement between the theoretical prediction and
observation.  If we flip the chirality of one of $\tilde{q}$ or
$\tilde{\tau}$, then the value of $\chi^2$ significantly increases; we
obtain $\langle\chi^2\rangle=15.8$ and $31.5$ for ${\cal L}=30$ and
$100\ {\rm fb}^{-1}$, respectively.  This provides important
information on the particles in the decay chain; the result indicates
that the squarks in the observed peak and the lighter stau have the
same chirality rather than different ones.  Information on the
chirality of $\tilde{\tau}$ may be obtained from other observables;
one of the examples is the tau polarization \cite{Kitano:2010tt}.
Then, assuming SUSY model as the underlying model, we obtain
information on the chirality of the dominantly produced squarks.

\begin{figure}[t]
  \centerline{\epsfxsize=0.75\textwidth\epsfbox{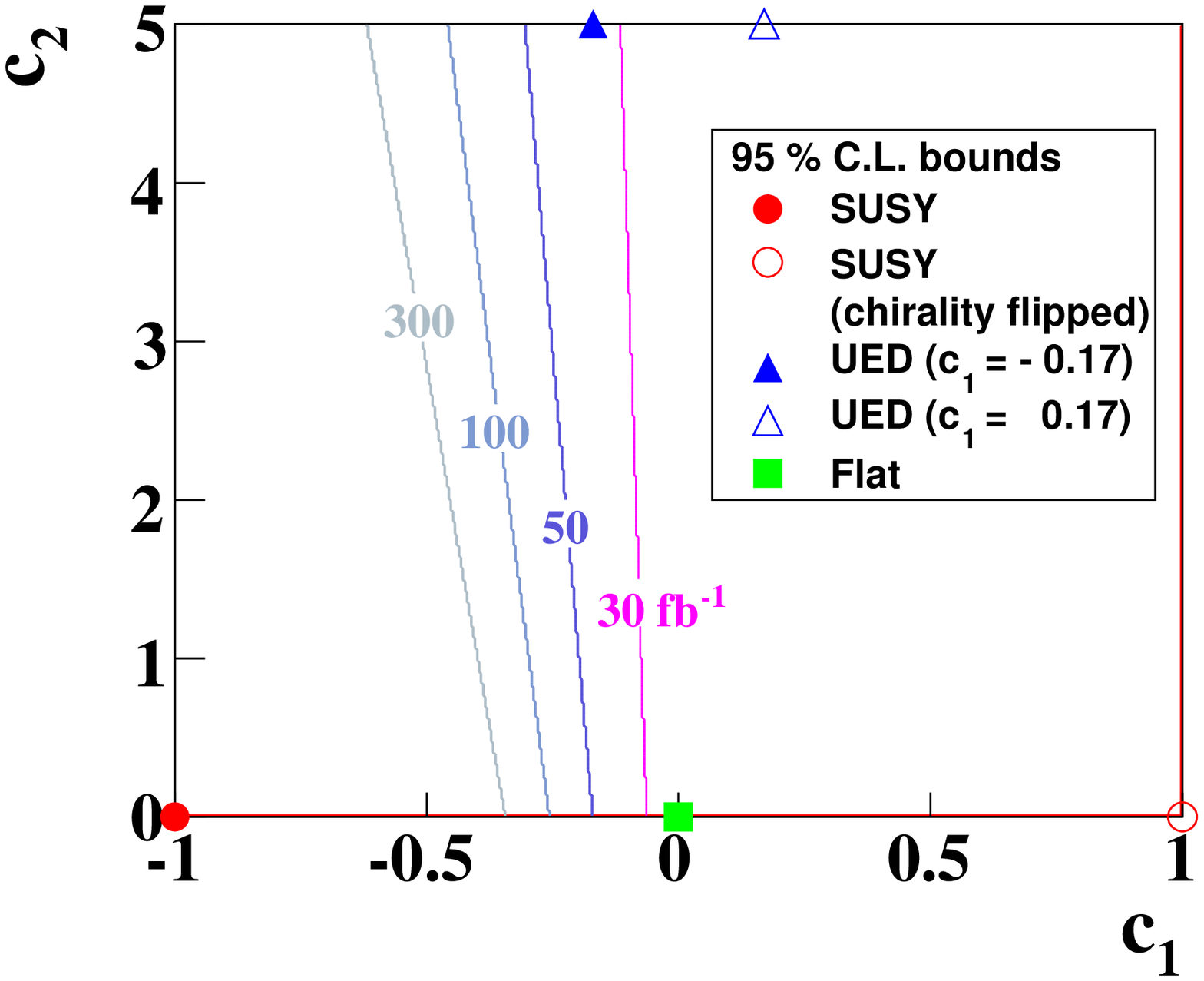}}
  \caption{\small 95 \% C.L. bounds on the $c_{1}$ vs.\ $c_{2}$ plane 
    for integrated luminosities indicated in the figure.  The region
    right to the contour is excluded for each luminosity.}
  \label{fig:chi2}
\end{figure}

Once the data from the SUSY events are collected in the LHC
experiment, we can also exclude class of models other than SUSY model
by studying the distribution of the invariant mass of $(q,\tau)$
system. For a model-independent analysis, we parameterize the
distribution of $x_{q\tau}$ for the decay chain resulting in
$\tau^\pm$ (and opposite-charge long-lived particle) as
\begin{eqnarray}
  \frac{1}{\Gamma_{q\tau^\pm}}
  \frac{d\Gamma_{q\tau^\pm}}{d x_{q\tau}}
  &=&
  \frac{6}{c_2+6}
  \left[ 1 \pm c_1 \left( 2 x_{q\tau} - 1 \right) 
    + c_2 x_{q\tau} (1 - x_{q\tau}) \right],
  \label{dG/dx(c1c2)}
\end{eqnarray}
where $c_1$ and $c_2$ are constants (with $|c_1|\leq 1$ and $c_2\geq
0$).  In the present SUSY model, $(c_1^{\rm (SUSY)}, c_2^{\rm
  (SUSY)})=(-1,0)$.  If the underlying model is assumed to be the UED
model, we would identify the particles with the masses of
$m_{\tilde{q}_R}$ and $m_{\chi^0_1}$ with KK modes of $q$ (denoted as
$q^{\rm (KK)}$) and neutral gauge boson (denoted as $B_\mu^{\rm
  (KK)}$), respectively, as well as the long-lived charged particle
with the KK mode of $\tau$ (denoted as $\tau^{\rm (KK)}$).  In this
framework, there exists a similar decay chain as in the present SUSY
model (i.e., $q^{\rm (KK)}\rightarrow qB_\mu^{\rm (KK)}$, followed by
$B_\mu^{\rm (KK)}\rightarrow\tau\tau^{\rm (KK)}$).  Then, we obtain
\begin{eqnarray}
  c_1^{\rm (UED)} &=& - \epsilon_q^{\rm (UED)} \epsilon_\tau^{\rm (UED)}
  \frac{2 m_{\chi^0_1}^4}{m_{\tilde{q}_R}^2 m_{\tilde{\tau}}^2 + 2 m_{\chi^0_1}^4},
  \\
  c_2^{\rm (UED)} &=& 
  \frac{4 m_{\chi^0_1}^2 \hat{M}_{q\tau}^2}
  {m_{\tilde{q}_R}^2 m_{\tilde{\tau}}^2 + 2 m_{\chi^0_1}^4},
\end{eqnarray}
where $\epsilon_f^{\rm (UED)}=+1$ when the fermion $f$ is right-handed
(i.e., singlet under $SU(2)_L$) while $\epsilon_f^{\rm (UED)}=-1$ when
$f$ is left-handed.  One can see that the distribution is different
from the SUSY case; this is due to the fact that the distribution
depends on the spins of particles in the decay chain.  Using the mass
spectrum of the present model, $(c_1^{\rm (UED)}, c_2^{\rm
  (UED)})\simeq (-0.17,5.0)$ or $(0.17,5.0)$, depending on the
relative chirality of the KK modes of $q$ and $\tau$.  (For
comparison, we also show the results for the UED case in Figs.\
\ref{fig:hist} and \ref{fig:ratio}, and Table \ref{table:Li} for the
case $(c_1^{\rm (UED)}, c_2^{\rm (UED)})\simeq (-0.17,5.0)$, which
results in a smaller value of $\chi^2$.)  Furthermore, in a model
where the invariant-mass distribution is flat in the phase space,
$(c_1^{\rm (flat)}, c_2^{\rm (flat)})=(0,0)$.  In Fig.\
\ref{fig:chi2}, we show the contours of $\langle\chi^2\rangle =14.1$,
which correspond to 95\ \% C.L. bounds for 7 degrees of freedom.  In
the same figure, we also show the points corresponding to various
models.  We can see that the analysis based on the ratio of the
numbers of $(j,j_\tau^+,\tilde{\tau}^-)$ and
$(j,j_\tau^-,\tilde{\tau}^+)$ events is useful for the test of
underlying models.  We note here that, on the contours given in Fig.\
\ref{fig:chi2}, $N_{\rm BG}$ is estimated to be smaller than the
number of signal events in average.  Thus, in the present case,
constraint on $c_1$ vs.\ $c_2$ plane can be obtained without knowing
the maximal possible number of backgrounds in detail; this is because
$L_i$ is strongly dependent on $x_{j\tau}$, as shown in Table
\ref{table:Li}, in the sample point used in our analysis.  In the case
with an underlying model giving rise to a weaker
$x_{j\tau}$-dependence of $L_i$, careful estimation of the maximal
number of backgrounds may be necessary to discriminate underlying
models.

In this letter, we have considered a procedure to study the properties
of superparticles in the case where the lighter stau is long-lived.
We have shown that properties of the particles in the decay chain are
extracted from invariant-mass distributions of the decay products of
$\tilde{q}$.  In particular, we emphasize that the effects of the
deformation of the invariant-mass distributions are largely reduced by
taking the ratio of two different decay processes with the same event
topology, which are, in the present case, $\tilde{q}\rightarrow
q\chi^0_1$, followed by $\chi^0_1\rightarrow\tau^+\tilde{\tau}^-$ and
by $\chi^0_1\rightarrow\tau^-\tilde{\tau}^+$.  We also note here that,
in some of the model beyond the standard model, a number of particles
(like neutralinos) decay into two different final states which are
charge-conjugated to each other, and that the reduction of the effects
of deformation by taking the ratio may be useful in various cases even
if there is no long-lived heavy charged particle.

In the present study, we have assumed that the squark(s) responsible
for the peak in the $M_{\tilde{q}}$ distribution has unique chirality.
This should be also experimentally confirmed.  One of the
circumstantial evidences of this may be negative observation of the
decay processes of squarks into Wino-like chargino and neutralino
(because the squarks are right-handed).  In addition, it may be also
possible to reconstruct $\tilde{q}_L$ production event to determine
the mass of the left-handed squarks, from which the observed peak in
the $M_{\tilde{q}}$ distribution may be understood to be from
$\tilde{q}_R$ production.  These will be discussed elsewhere
\cite{InPreparation}.

\noindent
{\it Acknowledgments:} 
The authors are grateful to R. Kitano for collaboration at an early
stage of this work.  They also thank M. Kakizaki for useful
discussion.  This work was supported in part by the Grant-in-Aid for
Scientific Research from the Ministry of Education, Science, Sports,
and Culture of Japan, Nos.\ 22540263 and 22244021 (T.M.).

\end{document}